\documentclass[aps,prb,preprint,groupedaddress,superscriptaddress]{revtex4}

\usepackage{graphicx}

\usepackage{dcolumn}
\usepackage{bm}

\begin{document}

\title{Amplification and squeezing of quantum noise with a tunable Josephson metamaterial}

\author{M.~A.~Castellanos-Beltran}
\email[Electronic mail: ]{castellm@colorado.edu} \affiliation{JILA,
National Institute of Standards and Technology and the University of
Colorado, and Department of Physics, University of Colorado,
Boulder, Colorado 80309}

\author{K.~D.~Irwin}
\author{G.~C.~Hilton}
\author{L.~R.~Vale}
\affiliation{National Institute of Standards and Technology,
Boulder, Colorado 80305}
\author{K.~W.~Lehnert}
\affiliation{JILA, National Institute of Standards and Technology
and the University of Colorado, and Department of Physics,
University of Colorado, Boulder, Colorado 80309}

\date{\today}

\maketitle

\textbf{It has recently become possible to encode the quantum state
of superconducting qubits and the position of nanomechanical
oscillators into the states of microwave fields\cite{Regal2008,
Houck}. However, to make an ideal measurement of the state of a
qubit, or to detect the position of a mechanical oscillator with
quantum-limited sensitivity requires an amplifier that adds no
noise. If an amplifier adds less than half a quantum of noise, it
can also squeeze the quantum noise of the electromagnetic vacuum.
Highly squeezed states of the vacuum serve as an important quantum
information resource. They can be used to generate entanglement or
to realize back-action-evading measurements of
position\cite{braunstein,Caves2}. Here we introduce a general
purpose parametric device, which operates in a frequency band
between 4 and 8 GHz. It is a subquantum-limited microwave amplifier,
it amplifies quantum noise above the added noise of commercial
amplifiers, and it squeezes quantum fluctuations by 10 dB.}

With the emergence of quantum information processing with electrical
circuits, there is a renewed interest in Josephson parametric
devices \cite{Castellanos,yurke2006,Haviland2007,Irfan,Nori}.
Previous work with Josephson parametric amplifiers demonstrated that
they can operate with subquantum-limited added noise and modestly
squeeze vacuum
noise\cite{yurke1987,yurke1988,yurke1989,yurke1990,yurke1996}.
Earlier realizations of Josephson parametric amplifiers (JPAs) were
only capable of amplifying signals in a narrow frequency range, were
not operated with large enough gain to make the noise of the
following, conventional amplifier negligible or were too lossy to be
subquantum limited\cite{Castellanos}. For related reasons, the
degree of squeezing of the vacuum noise was never larger than 3 dB.
We create a new type of parametric amplifier in which we embed a
tunable, low-loss, and nonlinear metamaterial in a microwave cavity.
The tunability of the metamaterial allows us to adjust the amplified
band between 4 and 8 GHz, and the cavity isolates the gain medium
from low-frequency noise, providing the stability required to
achieve high gains and large squeezing.

A single mode of a microwave field with angular frequency $\omega$
can be decomposed in two orthogonal components, referred to as
quadratures
\begin{displaymath}
\hat{V}(t)\propto \hat{X}_1\cos{\omega t} + \hat{X}_2\sin{\omega t}
\end{displaymath}
where $\hat{X}_1$ and $\hat{X}_2$ are conjugate quantum variables
obeying the commutation relation $[\hat{X}_1,\hat{X}_2]=i/2$. The
proportionality constant depends on the details of the
mode\cite{yurke1984,Louisell,Knight}. As a consequence of the
commutation relation, the uncertainties in $\hat{X}_1$ and
$\hat{X}_2$ are subject to the Heisenberg constraint $\Delta X_1
\Delta X_2 \geq 1/4$, where $\Delta X_j^2$ is the variance of the
quadrature amplitude $\hat{X}_j$. A mode is ``squeezed'' if for one
of the quadratures $\Delta X_j<1/2$ (ref. \onlinecite{Knight}). An
amplifier that transforms both input quadratures by multiplying them
by a gain $\sqrt{G}$ must add at least half a quantum of noise for
the output signal to obey the commutation relation\cite{caves1}; if
it adds exactly half a quantum of noise, it is quantum limited. On
the other hand, an amplifier which transforms the input signal by
multiplying one quadrature by $\sqrt{G}$ and multiplying the other
quadrature by $1/\sqrt{G}$ would have an output that satisfies the
commutation relation automatically. It need not add any noise to the
amplified output mode and, in that sense, can be subquantum
limited\cite{caves1,takahasi1965}. A degenerate parametric amplifier
transforms the input signal in this way, amplifying one quadrature
while deamplifying the other\cite{takahasi1965}.

\begin{figure}
\includegraphics{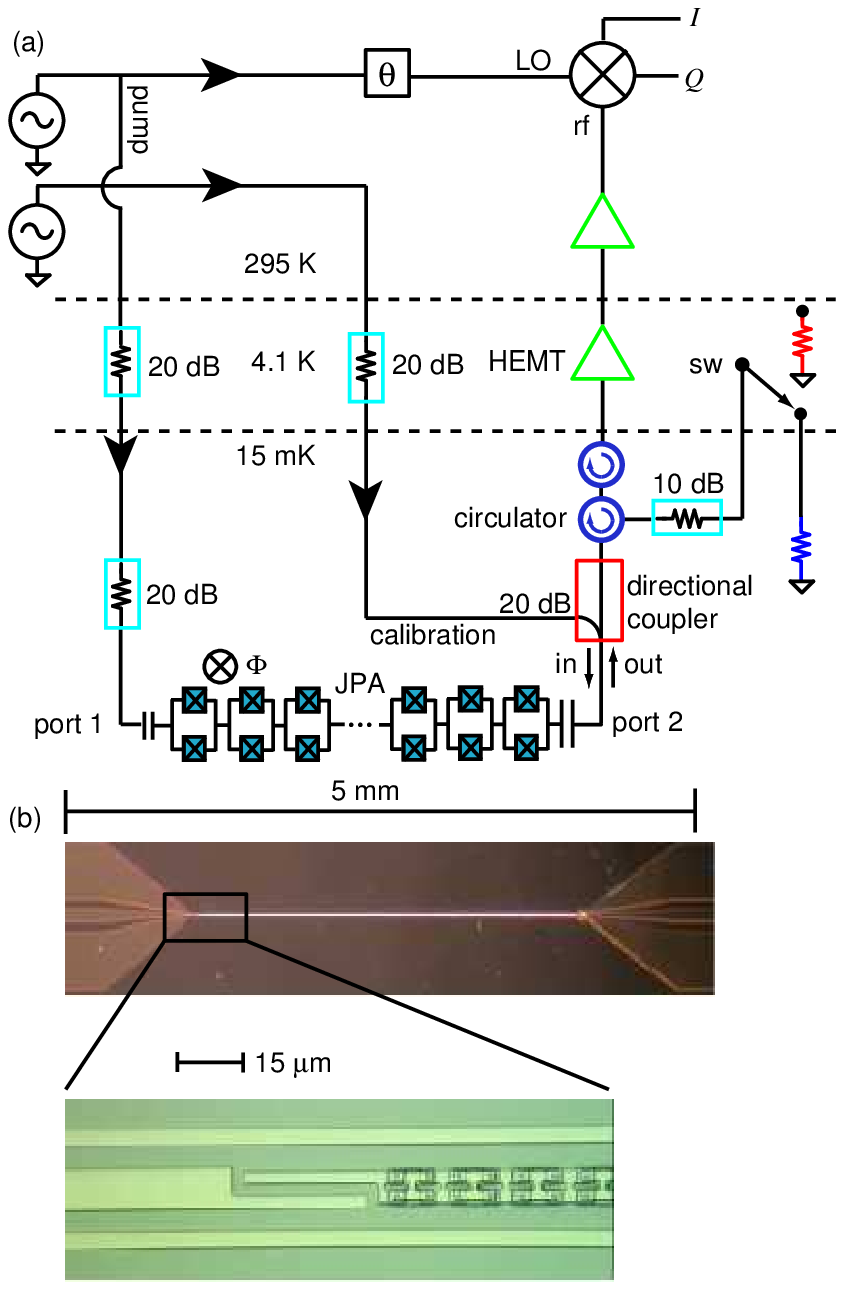}
\caption{\label{fig:Fig0}  \textbf{Measurement schematic and device
images. a,} The JPA is measured in a cryostat held at 15~mK. Two
microwave generators are used to study the JPA. One generator
creates the pump and LO tones, while the second creates a
calibration tone. The pump tone is injected through port 1 of the
JPA, while the calibration tone is incident on port 2 of the JPA
through a~20 dB direction coupler. Signals emerging from port 2 pass
through a circulator and are then amplified by a cryogenic
high-electron-mobility transistor amplifier (HEMT) and a room
temperature amplifier before entering the radio-frequency (rf) port
of a mixer. The noise incident on port 2 of the JPA which comes
primarily from the isolated port of the circulator can be switched
(sw) to come from two different resistors held at different
temperatures. \textbf{b,} The images are a picture of the full
device (upper) and a magnified image of the weakly coupled port and
a few SQUIDS (lower).}
\end{figure}

Our realization of the parametric amplifier consists of a
metamaterial embedded in a half-wavelength microwave cavity. The
metamaterial is a superconducting niobium coplanar waveguide in
which the center conductor is a series array of 480 Josephson
junctions. Each junction has been split into two junctions in
parallel, forming a SQUID and making the phase velocity of the
metamaterial tunable with magnetic flux enclosed by the SQUID
loop\cite{Haviland1996} (Fig. \ref{fig:Fig0}a). In addition to being
flux-tunable, the phase velocity depends on the intensity of the
fields propagating in this nonlinear metamaterial\cite{yurke1996}.
(The tunability and nonlinearity of this metamaterial were studied
in a preliminary device, fabricated from aluminum, which was too
lossy to be subquantum limited or to create highly squeezed states
\cite{Castellanos}.) We define the cavity by interrupting the center
conductor with two capacitors. One capacitor is larger than the
other, creating a strongly coupled port and a weakly coupled port.

We operate the parametric amplifier by injecting an intense pump
tone into the weakly coupled port at a frequency close to the
cavity's half-wave resonance. Through the intensity-dependant phase
velocity, the pump causes the cavity's resonance frequency to vary
at twice the pump's frequency providing parametric gain. A signal
incident on the strongly coupled port is reflected and will be
amplified if it is in phase with the pump but deamplified if it is
90 degrees out of phase with the pump. Signals exiting port 2 are
further amplified by a chain of commercial amplifiers. We resolve
the components of the amplified signal that are in phase $I$ and 90
degrees out of phase $Q$ by mixing the signal with a phase reference
called the local oscillator (LO). Because the LO is derived from the
same generator that produces the pump, $I$ and $Q$ are simply
related to the quadrature components $\hat{X}_1$ and $\hat{X}_2$ of
the microwave field exiting port 2. They are related by $ I =
A(\hat{X}_1\cos{\theta}-\hat{X}_2\sin{\theta} + \xi_I(t))$ and $ Q =
A(\hat{X}_1\sin{\theta}+\hat{X}_2\cos{\theta} + \xi_Q(t))$, where
$A$ is the total gain of the commercial amplifiers and mixer,
$\theta$ is the phase between the pump and the LO, and $\xi_I$ and
$\xi_Q$ are random variables. Both random variables have a power
spectral density $N_A$, where $N_A$ is the noise number\cite{caves1}
of the commercial amplifier chain expressed as noise added at the
input of the HEMT (see Methods).

\begin{figure}
\includegraphics{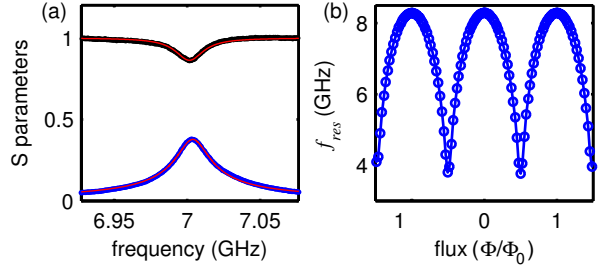}
\caption{\label{fig:Fig1}  \textbf{Linear response of the JPA. a,}
We plot the fraction of microwave amplitude transmitted from port 1
to port 2 $|\mathrm{S_{21}}|$ (blue) and the fraction reflected from
port 2 $\mathrm{|S_{22}}|$ (black) as functions of frequency at a
particular value of applied magnetic flux. By fitting this data to a
model of a two port cavity (red lines), we extract $f_{res}=7.0038
$~GHz, $\gamma_{c1}=2\pi \times 437$~kHz, $\gamma_{c2}=2\pi \times
10.55$~MHz, and $\gamma_{i}=2\pi \times 341$~kHz. \textbf{b,} We
plot the measured value of $f_{res}$ as a function of applied
magnetic flux $\Phi$ in units of the magnetic flux quantum $\Phi_0$.
We use the obvious periodicity in this data to infer the magnetic
flux enclosed by the SQUIDs.}
\end{figure}

Before operating the device as a parametric amplifier, we first
characterize its linear behaviour. In Fig. \ref{fig:Fig1}a, we
measure the cavity's response both by measuring the reflectance of a
signal on port 2 and the transmittance from port 1 to port 2 as
functions of frequency. The powers used are low enough so that the
cavity's response is still linear. From this measurement we can
extract the cavity's resonance frequency $f_{res}$, the rates
$\gamma_{c1}$ and $\gamma_{c2}$ at which the cavity loses power
through both ports as well as through internal loss processes
$\gamma_{i}$ (Fig.~\ref{fig:Fig1}a). When operating as a parametric
amplifier, the center of the amplified band will be close to
$f_{res}$, and the unity gain signal bandwidth will be approximately
$\gamma_{c1}+\gamma_{c2}+\gamma_{i}$\cite{yurke2006,Castellanos}. By
applying magnetic flux we can move the center of the amplified band
over a large range of frequencies (Fig. \ref{fig:Fig1}b).

\begin{figure}
\includegraphics{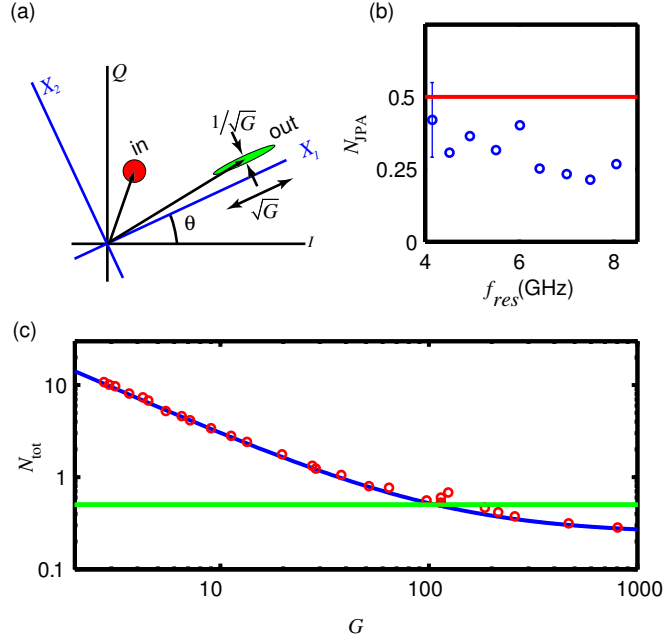}
\caption{\label{fig:Fig2}\textbf{ Added noise of the JPA and of the
full measurement chain. a,} This diagram represents the operation of
an ideal degenerate parametric amplifier which transforms an input
state (red circle) into an output state (green ellipse) by
amplifying the $X_1$ component by $\sqrt{G}$ and squeezing the $X_2$
component by $1/\sqrt{G}$. The mixer projects the output state into
axes rotated by $\theta$. \textbf{b,} We measure $N_{\mathrm{JPA}}$
at nine different values of $f_{res}$ between 4 and 8 GHz and
observe that the added noise is less than half a quantum (red line).
The error bar applies to all points; it represents the systematic
uncertainty introduced primarily by the unknown loss of the
directional coupler and switch when operated cryogenically.
\textbf{c,} We plot $N_{\mathrm{tot}}$ when $f_{res}= 7.00$ GHz
versus $G$ (points). At this frequency $N_A=26$ and is dominated by
the HEMT. With increasing $G$, this noise becomes negligible
compared to $N_{\mathrm{JPA}}=0.23$, where we expect
$N_{\mathrm{tot}}=N_{\mathrm{JPA}}+N_A/G$ (blue line). At the
largest gains $N_{\mathrm{tot}}<0.5$ (green line) and $N_A/G=0.04$.}
\end{figure}

After characterizing the linear response of the device, we operated
it as an amplifier. In previous work, we studied the dependence of
the parametric gain on the pump power and pump frequency in our
first realization of this metamaterial\cite{Castellanos}, finding
good agreement with the theory in ref.~\onlinecite{yurke2006}. Here
we study the JPA's added noise as a function of the cavity's
resonance frequency and the total noise of the amplifier chain as a
function of the JPA gain. By switching between two calibrated noise
sources (Fig. \ref{fig:Fig0}a), we can change the incident noise on
the JPA. The change in noise power at the output of the mixer allows
us to measure the noise $N_{\mathrm{JPA}}$ added by the JPA, the JPA
gain $\sqrt{G}$, the total added noise of both the JPA and the
commercial amplifier chain $N_{\mathrm{tot}}$ (i.e., the total noise
referred to the input of the JPA), and the total gain $\sqrt{G}A$.
The calibrated noise sources are derived from two resistors held at
different temperatures. The resistances are matched to the wave
impedance of the cables that carry signals to the JPA input. The
noise emitted by both resistors is attenuated by a 10~dB attenuator
held at a temperature $T_A=15$~mK and then passed to the JPA through
a microwave circulator which separates the incident and reflected
signals from the JPA. The power spectral density in units of noise
quanta incident on the JPA through the circulator is
\begin{equation}
    N(T_R)=\frac{1}{2}+\left(\frac{9}{10}\right)\frac{1}{e^{\frac{\hbar\omega}{k_BT_A}}-1}+\left(\frac{1}{10}\right)\frac{1}{e^{\frac{\hbar\omega}{k_BT_R}}-1},
    \label{eq:Nt}
\end{equation}
where $T_R=T_c=15$~mK or $T_R=T_h=4.1$~K depending on the position
of the switch shown in Fig. \ref{fig:Fig0}a. The first term in
equation~(\ref{eq:Nt}) is the quantum noise while the second and
third terms are the thermal noise which is small compared to the
quantum noise at $T_R=T_c$. Most of the noise incident on the JPA is
$N(T_R)$, but a small amount comes from the coupled port of the
directional coupler (see Methods). By selecting $\theta=0$, the
amplified quadrature appears at the $I$ port of the mixer (Fig.
\ref{fig:Fig2}a). We measure the ratio $Y$ of the noise power at the
output $I$ with the hot load incident on the JPA divided by the
noise power with the cold load incident on the JPA. This
measurement, known as a $Y$-factor measurement, allows us to extract
$N_{\mathrm{tot}}=(N(T_h)-YN(T_c))/(Y-1)$ and the total gain
$A\sqrt{G}$. We then adjust the cavity's resonance frequency about
10 linewidths away from the LO frequency and turn off the pump.
Because the JPA now acts as a passive mirror, we can perform a
$Y$-factor measurement on our commercial amplifier chain, finding
$N_A$ and $A$. From both measurements, we extract the JPA gain
$\sqrt{G}$ and $N_{\mathrm{JPA}}$ (Fig.  \ref{fig:Fig2}). As
observed in Fig. \ref{fig:Fig2}b, our JPA adds less than 1/2 a
quantum of noise over the $4 - 8$ GHz band where it operates. In
addition, it can be operated at large enough gain to amplify the
quantum noise above the commercial amplifier's noise (Fig.
\ref{fig:Fig2}c)!

\begin{figure}
\includegraphics{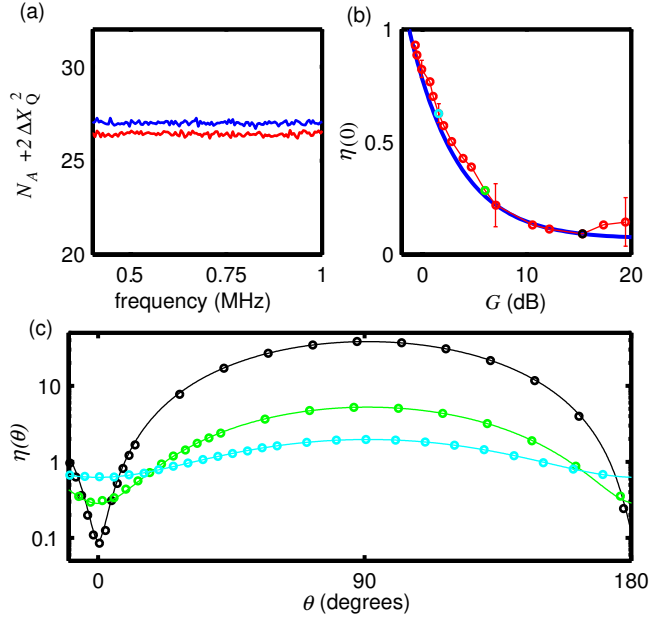}
\caption{\label{fig:Fig3} \textbf{Demonstration of squeezing. a,}
Power spectral density versus frequency at the $Q$ output of the
mixer referred to the input of the commercial amplifier chain in
units of quanta, measured with the pump on (red) and the pump off
(blue) for $\theta=0$, $f_{res}=7$~GHz and $G=15.3$~dB. \textbf{b,}
The squeezing $\eta(0)$ as a function of $G$ (points) and the
predicted squeezing from reference \onlinecite{yurke2006} (line).
The error bars represent the same systematic error as in
Fig.~\ref{fig:Fig2}b. This uncertainty varies smoothly between the
error bars shown in selected data points. \textbf{c,} $\eta(\theta)$
as a function of $\theta$ for three different gains (points) for
$G=34$ (black), $G=4$ (green), $G=1.4$ (cyan) and the expected
$\theta$ dependance (lines)\cite{yurke1989,yurke2006}.}
\end{figure}

In order to show that the JPA can squeeze quantum noise, we examine
the squeezed quadrature $X_2$ when the incident noise is mostly
quantum noise ($T_R=T_c$). When $\theta=0$, $X_2$ appears at the $Q$
port. The noise at the $Q$ port referred to the input of the HEMT is
composed of the added noise of the commercial amplifiers
($N_A\approx26$) and noise coming from port 2, $2\Delta X_Q^2$,
where $\Delta X_Q^2=\Delta X_1^2\sin^2{\theta}+\Delta
X_2^2\cos^2{\theta}$. When $\theta=0$, we observe a reduction in
$\Delta X_Q$ when the pump is turned on, demonstrating that the JPA
has squeezed quantum noise (Fig. \ref{fig:Fig3}a). As we increase
the JPA gain, we observe an increase in the noise squeezing up to
$G=16$~dB (Fig.~\ref{fig:Fig3}b). For $G<16$~dB the squeezing we
observe is consistent with the expected behaviour from a model of
parametric amplification that includes loss in the JPA but no other
imperfections\cite{yurke2006}. However, for $G>16$~dB we observe
less than the predicted squeezing probably due to the instability in
the phase acquired by the pump as it passes through the cryostat in
the several minutes required to complete one measurement. Finally,
to illustrate the JPA's unequal effect on the noise of the two
quadratures, we define a quantity $\eta(\theta)=(\Delta
X_Q^2)_\mathrm{pump\ on}/(\Delta X_Q^2)_\mathrm{pump\ off}$. In
Fig.~\ref{fig:Fig3}c, we plot $\eta(\theta)$ demonstrating that the
fluctuations at the output of the JPA are indeed squeezed along the
axis defined by the pump (Fig.~\ref{fig:Fig2}a).

The measurement of the squeezed state would have proceeded much more
rapidly if we had used a subquantum limited amplifier, such as a
second JPA, instead of a HEMT amplifier. For most measurements, the
low added noise of the JPA is helpful, but for some measurements it
is crucial. For example, to fully characterize the quantum state of
non-Gaussian states, such as the Fock states generated by
superconducting qubits \cite{Houck}, a subquantum limited amplifier
is a necessity\cite{Stoler}. In addition to characterizing
non-Gaussian states of microwave fields, the JPA is also well suited
to amplifying the signal generated by a nanomechanical beam moving
in a microwave cavity\cite{Regal2008}, enabling a quantum-limited
measurement of position. The tunability of the amplifier
demonstrated here also makes it well suited to use with the
increasingly popular frequency-division-multiplexed microwave
circuits\cite{Regal2008, Day, Mates} when simultaneous measurement
of all frequency channels is not required.

\section*{Methods}

Devices are fabricated at NIST Boulder using a standard Nb/AlOx/Nb
trilayer process \cite{NISTfab}, modified by eliminating the shunt
resistor layer and minimizing deposited oxides \cite{Mates}. The
device studied here was coprocessed with 10 other devices used for
dc characterization. These dc-measurements indicate an average
critical current $I_c$ per SQUID in the JPA of 31~$\mu$A, which was
close to the designed value of 30~$\mu$A. The transmission lines
from which the half-wavelength cavities were built were designed to
have a capacitance per unit length of $C_l=0.15$ nF/m and a
geometrical inductance per unit length of $L_l=0.49$ $\mu$H/m. For
$I_c=31\ \mu$A, the Josephson inductance per SQUID is $L_J=10.6$~pH.
These values predicts a half-wave resonance frequency of $8.23$ GHz,
when no flux is applied, which is within 3\% of the observed value.
Measurements were performed in a magnetically shielded dilution
refrigerator. To avoid unknown loss and noise in the cables carrying
the calibrate noise signals, superconducting Niobium coaxial cables
were used to carry these signals between the 15 mK region of the
refrigerator and 4~K region. A calibrated ruthenium oxide
thermometer measured the refrigerators base temperature and the
temperature of the helium bath, to which the 4~K resistor and
microwave switch were thermally anchored. A calibration tone, (Fig.
1a) is used: to verify the gain inferred from the $Y$-factor
measurements, to ensure that the noise reduction arises from genuine
squeezing and not saturation of the amplifiers, and to measure
$\mathrm{S_{22}}$ in the network analysis. The complete expression
for the noise incident on the JPA includes contributions from the
thermal noise introduced with the calibration tone and the insertion
loss of the directional coupler and the switch. The unknown
insertion loss at cryogenic temperatures of the directional coupler
and of the microwave switch are the dominant uncertainties in the
added noise and squeezing measurements. The extreme values
represented by the error bars are derived by assuming: first, that
there in no insertion loss, and then, that the insertion loss has
not changed between room temperature and the operating temperature.
Noise at the $I$ or the $Q$ ports is expressed as the
double-side-band power-spectral-density referred to the input of the
relevant amplifier. Dividing the spectral density by
$hf_{\mathrm{LO}}$, yields the noise in dimensionless units of noise
quanta, where LO is the local oscillator frequency; e.g. the thermal
noise of a resistor at temperature $T>>h f_{LO}/k_B$ measured at the
$I$-port would have a noise $N_T=k_BT/hf_{\mathrm{LO}}$.

\section*{Acknowledgements}

The authors acknowledge funding from the National Institute of
Standards and Technology (NIST), from the National Science
Foundation, and from a NIST-University of Colorado seed grant. We
thank S.~M. Girvin, J. D. Teufel, C. A. Regal, N. E. Flowers-Jacobs,
J. K. Thompson and M. Holland for valuable conversations and
technical assistance. K.~W. Lehnert is a member of NIST's Quantum
Physics Division.

\section*{Competing financial interests}

The authors declare that they have no competing financial interests.


\end{document}